\documentclass[referee]{raa}           
\usepackage{graphicx,times}
\usepackage{natbib}
\usepackage{amssymb,amsmath}
\newcommand{\Ion}[2]{#1{\,\scriptsize #2}}
\bibpunct{(}{)}{;}{a}{}{,}

\usepackage[a4paper=true,pagebackref=true]{hyperref}
\hypersetup{pdftitle = The title of my PDF, pdfauthor = My name, pdfsubject= The subject, pdfkeywords = keyword1 keyword2 keyword3} 
\hypersetup{colorlinks = true, linkcolor = green, anchorcolor = red, citecolor = blue, filecolor = red, pagecolor = red, urlcolor = red}

\begin{document}

   \title{A catalogue of early-type emission-line stars and H$\alpha$ line profiles from LAMOST DR2
$^*$
\footnotetext{\small $*$ Supported by the National Key Basic Research Program of China(973) and the National Natural Science Foundation of China(NSFC).}
}

 \volnopage{ {\bf 2012} Vol.\ {\bf X} No. {\bf XX}, 000--000}
   \setcounter{page}{1}

   \author{Wen Hou\inst{1,2}, ALi Luo\inst{1,2}, Jingyao Hu\inst{1}, Haifeng Yang\inst{1,2,3}, Changde Du\inst{1,2}, Chao Liu\inst{1}, Chien-De Lee\inst{4} , Chien-Cheng Lin\inst{5}, Yuefei Wang\inst{6},Yong Zhang\inst{6}, Zihuang Cao\inst{1}, and Yonghui Hou\inst{6}
   }

   \institute{ Key Laboratory of Optical Astronomy, National Astronomical Observatories, Chinese Academy of Sciences,
Beijing 100012, China; {\it lal@bao.ac.cn, whou@bao.ac.cn}\\
        \and
             University of Chinese Academy of Sciences, Beijing 100049, China\\
\and 
School of Computer Science and Technology, Taiyuan University of Science and Technology, Taiyuan,030024, China\\
\and
Institute of Astronomy, National Central University, Jhongli, Taiwan
\and
Shanghai Astronomical Observatory, Chinese Academy of Sciences, Shanghai 200030, China\\
\and
Nanjing Institute of Astronomical Optics $\&$ Technology, National Astronomical Observatories, Chinese Academy of Sciences, Nanjing 210042, China
\vs \no
   {\small Received  ; accepted  }
}

\abstract{ We present a catalogue including 11,204 spectra for 10,436 early--type emission--line stars from LAMOST DR2, among which 9,752 early--type emission--line spectra are newly discovered. For these early--type emission--line stars, we discuss the morphological and physical properties from their low-resolution spectra. In this spectral sample, the H$\alpha$ emission profiles display a wide variety of shapes. Based on the H$\alpha$ line profiles, these spectra are categorized into five distinct classes: single--peak emission, single--peak emission in absorption, double--peak emission, double--peak emission in absorption, and P--Cygni profiles. To better understand what causes the H$\alpha$ line profiles, we divide these objects into four types from the view of physical classification, which include classical Be stars, Herbig Ae/Be stars, close binaries and spectra contaminated by \Ion{H}{II} regions. The majority of Herbig Ae/Be stars and classical Be stars are identified and separated using the ($\emph{H}$-$\emph{K}$, $\emph{K}$-$\emph{W1}$) color--color diagram. We also discuss thirty one binary systems as listed in SIMBAD on--line catalogue and identify 3,600 spectra contaminated by \Ion{H}{II} regions after cross–matching with positions in the \citeauthor{1976A&AS...25...25D} catalogue. A statistical analysis
of line profiles versus classifications is then conducted in order to understand
 the distribution of H$\alpha$ profiles for each type in our sample. Finally, we also provide a table of 172 spectra with \Ion{Fe}{II} emission lines and roughly calculate stellar wind velocities for seven spectra with P--Cygni profiles.
}

   \authorrunning{W., Hou, A-li Luo et al. }            %author_head in even pages
   \titlerunning{Early--type emission--line stars from LAMOST DR2}  % title_head in odd pages
   \maketitle

%________________________________________________ sections below
%
\section{Introduction}           %% first-level sections will be auto-capitalized
\label{sect:intro}
The strong optical emission lines which characterize hot emission-line stars enable the study of stellar envelopes, often revealing accretion flows, stellar winds and binary interaction. The objects with emission lines are widely distributed on the Hertzsprung-Russell diagram including various stellar types. \citet{2007ASSL..342.....K} divided the emission-line stars into four types, which are respectively early-type stars (Of, Oe/Be/Ae, etc), late type stars (dMe, Mira variables, etc), close binaries (Algol stars, cataclysmic variables and symbiotic stars), and Pre-main sequence stars (Herbig Be/Ae stars and T Tauri stars). It is clear that the H$\alpha$ emission line may be present in any of the early--type spectral classes. In general, the H$\alpha$ emission stars with O, B or A spectral type are dominated by three types which are classical Be (Oe/Ae) stars (hereafter referred to CBe stars), Herbig Ae/Be stars (hereafter HAeBe stars) and close binaries.

CBe stars are characterized by emission in the Balmer lines, sometimes accompanied by emission in lines of singly-ionized metals or neutral helium \citep{2007ASSL..342.....K,2009ssc..book.....G}. \citet{1981BeSN....4....9J} gave a more precise definition confining them to the stars of luminosity class V-III. The comprehensive reviews of understanding CBe stars have been given in several works \citep{1988PASP..100..770S,2003PASP..115.1153P,2013A&ARv..21...69R}. Different from CBe stars, HAeBe stars are pre-main sequence stars with a mass of 2$\sim$10 solar mass \citep{1998ARA&A..36..233W}. Besides the emission lines of Balmer series and some ionized metals such as \Ion{Fe}{II}, which exist in spectra of CBe stars, low-excitation lines such as \Ion{Ca}{II} and \Ion{Fe}{I} also can be shown in the spectra of HAeBe stars. Properties that distinguish HAeBe from CBe stars are discussed in detail by \citet{1960ApJS....4..337H} and \citet{1972ApJ...173..353S}. Many studies have also been focused on the IR color analysis of these two types, and distinguish them based on the IR excess in recent decade \citep{1974ApJ...191..675G, 1984A&AS...55..109F, 1987A&A...176...93C, 1990AcApS..10..154H, 2006Ap&SS.305...11Z,2015RAA....15.1325L}. In addition to CBe and HAeBe stars, close binaries exhibiting both the H$\alpha$ emission line and an early--type spectrum are usually comprised of Algol--type systems and cataclysmic variable stars such as nova\citep{2007ASSL..342.....K}. %An Algol--type binary, is a interacting binary system where most of the primaries are B or early A type main-sequence stars and the companions are G or K giants or subgiants fulfilling the Roche lobe\citep{2007ASSL..342.....K}. Cataclysmic variable stars, mainly nova or nova--like stars, which are a type of stars consisting of a close binary star system in which one of the components is a white dwarf that accretes matter from its companion.

It has been recognized that H$\alpha$ emission line in the spectra originates from stellar envelopes or outer stellar atmospheres in early--type stars, due to the presence of strong stellar winds, rotating rings and accretion dust disks. Moreover, depending on the different physical mechanisms or the geometry of the objects, the H$\alpha$ emission profiles display a wide variety of shapes. Many attempts have already been made to investigate  their physical origin, the H$\alpha$ profile features, and the classification of line profiles \citep{1979ApJ...230L..99U, 1984A&AS...55..109F, 1988A&A...189..147H, 1996A&AS..120..229R,2012MNRAS.425..355R, 2013ApJ...765...41S, 2013A&A...556A..81B, 2013A&A...556A.108B, 2015A&A...581A..52T}.

From the previous studies, almost all the morphological classifications of H$\alpha$ profiles are based on small samples of high--resolution spectra. In order to conduct a strong statistical analysis we employ a total of ~200,000 early-type stellar spectra acquired by the Large sky Area Multi-Object Spectroscopic Telescope (LAMOST) survey up until June, 2014. From these observations we select a sample of 11,204 early--type emission-line spectra, representing 10,436 unique stars.  For these early--type H$\alpha$ emission stars, we have made analyses from two aspects of the H$\alpha$ line profiles and physical classification respectively. In section 2, we briefly describe the data and sample selection. A spectral analysis based on the H$\alpha$ line profiles is presented in section 3. We also give a detailed description of six morphological categories with additional unclassified profiles. In section 4, a physical classification of these early--type H$\alpha$ emission stars is provided.  In section 5, we make an analysis of two particular classes: spectra with P--Cygni profiles and metal emission lines. Finally, a brief summary is provided in section 6.

%\section{}
%\section{data and sample selection}
\section{data and sample selection}
\subsection{The LAMOST data}
The LAMOST with an effective aperture of 4 m and a large field of view of 5$^{\circ}$, is equipped with 16 spectrographs, 32 CCDs and 4,000 fibers. A spectroscopic survey launched by this large instrument started in October 2011. It is planed to cover approximately half of the celestial sphere, and will eventually collect 10 million spectra of stars, galaxies and QSOs with the resolution of $\sim$1800 and the wavelength coverage of 3650---9000 \AA. By June 2014, the first--year pilot survey and first two years of regular survey, more than 4 million spectra are observed, among which there are about 3.78 million stellar spectra. In the large dataset, $\sim$ 200,000 spectra which occupy about 5\% of the stars in DR2 are classified as as O, B and A--type by LAMOST 1D pipeline \citep{2015RAA....15.1095L,2016luoal}. A total of 200,259 O, B and A type stars for H$\alpha$ emission detection are collected from DR2, among which the majority are A type stars.

\subsection{Detection of H$\alpha$ emission line}
In order to pick out early--type stars with H$\alpha$ in emission, we propose a subjective criteria for detecting H$\alpha$ line profiles. A brief description of the criteria is given here. For all the spectra of O, B and A--type stars, we use Eq. \ref{detection1} as the first criterion to pick out the spectra with H$\alpha$ line in emission. This equation indicates that the average flux over the wavelength within five pixels of the H$\alpha$ line center is larger than the continuum flux. It is useful for the H$\alpha$ emission lines with ordinary intensity above the continuum.
\begin{equation} \label{detection1}
 \sum_{i=-5}^{5}f_{obs}[n_{0}~+~i]~/~11 ~>~ f_{conti}[n_{0}]
\end{equation}
where n$_{0}$ is the pixel index of central wavelength of H$\alpha$ ($\lambda_{0}$ = 6564 \AA), and $f_{obs}$, $f_{conti}$ denote the flux for observed spectra and continuum spectra respectively.
A sample of twenty thousand spectra is randomly selected from $\sim$ 200,000 O, B, and A stars for Eq. \ref{detection1} testing. Our manual checks confirmed that almost all the spectra with strong H$\alpha$ emission lines can be picked out by use of this equation, except for some spectra with a weak emission lying in the deep H$\alpha$ absorption profiles. To retrieve such spectra, we provide Eq. \ref{detection2} which is engaged in a narrow feature of H$\alpha$ with fewer pixels, for detecting spectra with low--intensity H$\alpha$ emission lines.
\begin{equation} \label{detection2}
 \begin{aligned}
 & \sum_{i=-1}^{1}f[n_{0}~+~i]~/~3 > \sum_{i=-2}^{2}f[n_{0}~+~i]~/~5 ~~~~~~~\& \\
 & max(f_{obs}[n_{0}-1:n_{0}+1]) \geqslant max(f_{obs}[n_{0}-2:n_{0}+2]) 
 \end{aligned}
\end{equation}
where the denotations of n$_{0}$, $f_{obs}$, $f_{conti}$ are the same as Eq. \ref{detection1}, and max(f$_{obs}$[x:y]) represents the maximum flux where the index is from x to y. 

\begin{figure}[!b]
   \centering
   \includegraphics[width=10.0cm, angle=0]{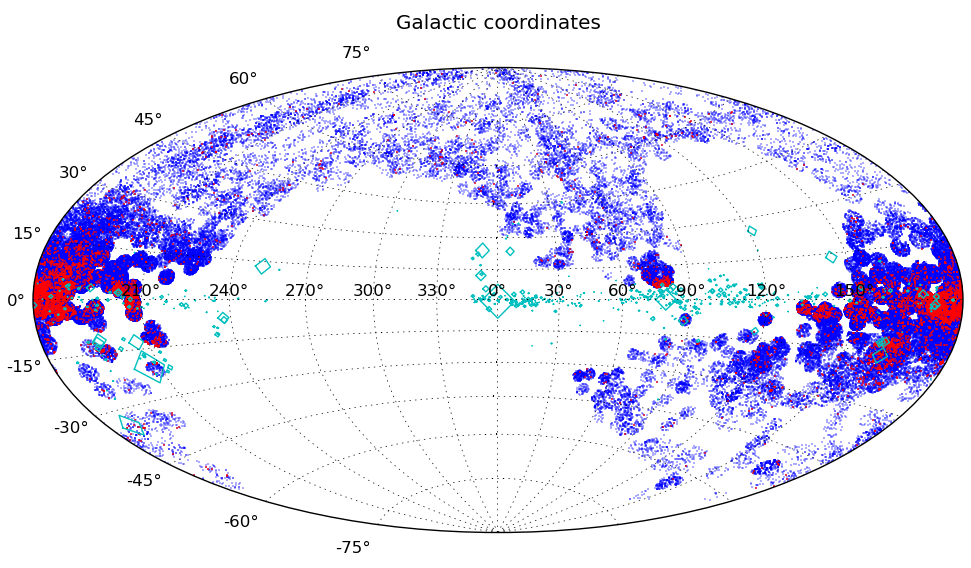}
  % \begin{minipage}[]{85mm}
   \caption{The spatial distribution of both the initial sample and early--type H$\alpha$ emission stars from LAMOST DR2. The blue and red points represent $\sim$ 200,000 O, B, A stars and 11,204 early--type H$\alpha$ emission-line stars respectively. These two samples are concentrated in the region of Galactic Anti--centre due to the observational strategy. The cyan boxes represent \Ion{H}{II} regions which are discussed in section 4.2.} 
%\end{minipage}
   \label{Figadd_1}
   \end{figure}

Applying these criterion to the initial sample, spectra satisfied either criterion are picked out.
It is noted that the noises of spectra may affect the selection of H$\alpha$ emission lines, especially for spectra with a weak H$\alpha$ emission line as well as a low signal to noise for r band (SN$\_$r $<$ 10). To be specific, the detections of H$\alpha$ emission lines are less accurate for the spectra with lower SN$\_$r. In order to ensure the results as accurately as possible, we visually inspect all the selected spectra one by one, abandoning spectra with H$\alpha$ profiles severely affected by noise. Finally, a total of 11,204 spectra with H$\alpha$ in emission are selected from LAMOST DR2. The spatial distribution of this sample in Galactic coordinates plotted by red points is shown in Figure \ref{Figadd_1}. We can see that this sample are concentrated in the region of Galactic Anti–centre due to the observational strategy. The total number of spectra with H$\alpha$ emission and the total number of stars for each spectral type are presented in Table \ref{table1}.

% figures.

\begin{table}
\bc
\begin{minipage}[]{160mm}
\caption[]{The number of spectra with H$\alpha$ in emission as well as the total spectra for O, B, and A type respectively. \label{table1}}\end{minipage}
\setlength{\tabcolsep}{8pt}
\small
 \begin{tabular}{c|cccc}
  \hline\noalign{\smallskip}
Spectral type & O type & B type & A type \\
  \hline\noalign{\smallskip}
Number of spectra & 14 & 492 & 10,698 \\
with H$\alpha$ emission lines \\
  \hline\noalign{\smallskip}
Number of  & 117 & 3,137 & 197,005 \\
the total spectra \\
  \noalign{\smallskip}\hline
\end{tabular}
\ec
%% place \tablecomments and \tablerefs below \end{center| and \end{center}:
%% you may leave the table-width parameter to editors or set to your actual size
\end{table}
 
\section{Morphological classification of H$\alpha$ profile}
\begin{figure}
   \centering
   \includegraphics[width=14.0cm, angle=0]{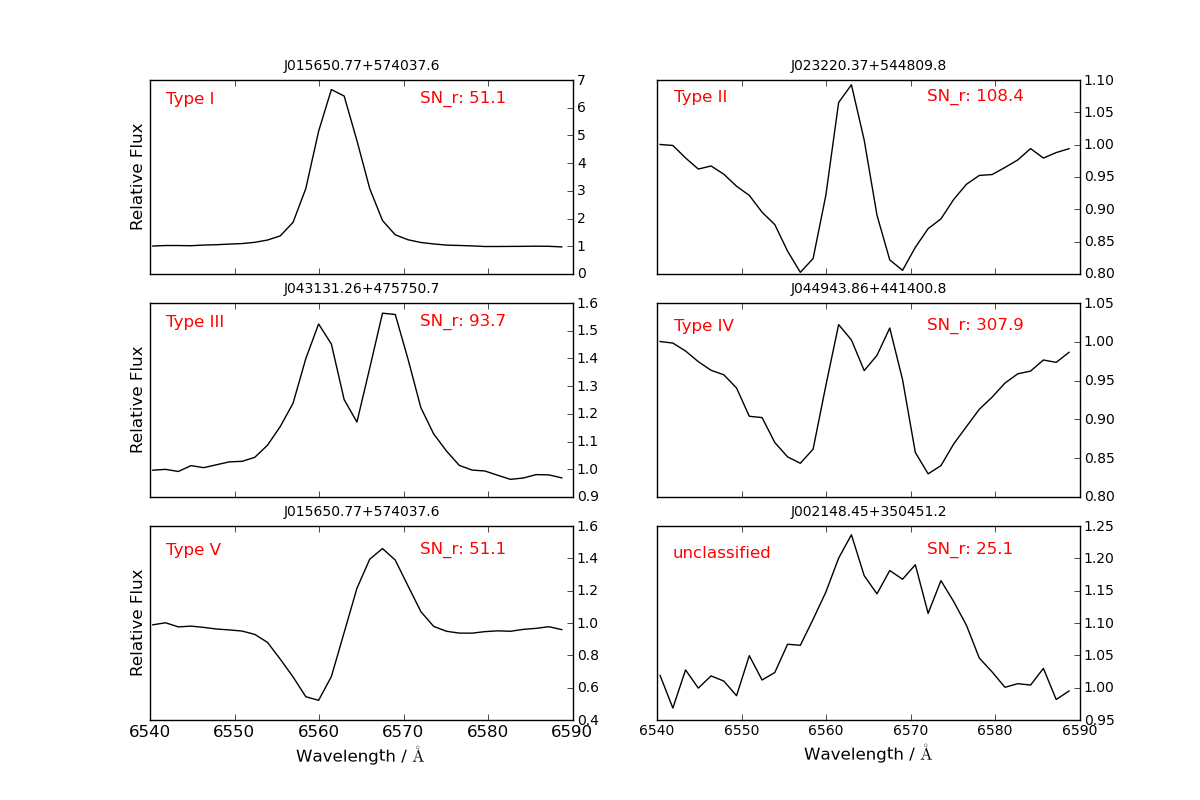}
  % \begin{minipage}[]{85mm}
   \caption{Examples of six spectra with H$\alpha$ emission line of different types. These spectra with SN$\_$r $>$ 50 (except the unclassified one) are observed by LAMOST instruments with the resolution of $\sim$1800. Radial velocities are not removed for these LAMOST spectra. } 
%\end{minipage}
   \label{Fig1}
   \end{figure}

We endeavour to make a morphological classification of H$\alpha$ emission profile for the sample of low--resolution spectra. Quite similar to the classification system provide by \cite{1984A&AS...55..109F}, here we divide the spectra in our sample into five morphological categories: emission with single peak, emission with single peak in absorption, emission with double peaks, emission with double peaks in absorption, and P--Cygni. The only difference between the first two categories is whether the emission lies in an apparent absorption component. So is the third and the fourth case. Five spectra (SN$\_$r $>$ 50) with H$\alpha$ emission profiles of basic types as well as one spectra with unclassified profile from LAMOST survey are shown in Figure \ref{Fig1}.

% figures.

When making the morphological classification, it is noted that the shape of H$\alpha$ profile is probably affected by the SN$\_$r of spectra. The classification results are more reliable for spectra with a high SN$\_$r and vice-versa. Therefore, to incorporate reliability into the analysis, the dataset of spectra with H$\alpha$ in emission from O, B and A--type stars in DR2 are divided into five groups based on SN$\_$r which can indicate noise level of the continuum around the H$\alpha$ feature. These five groups are respectively subsamples with SN$\_$r $<$ 5, 5 $\leq$ SN$\_$r $<$ 10, 10 $\leq$ SN$\_$r $<$ 30, 30 $\leq$ SN$\_$r $<$ 50 and SN$\_$r $\geq$ 50. %%This paragraph introduce the influence of SN$\_$r on the classification of Ha profile.

\begin{figure}
   \centering
   \includegraphics[width=12.0cm, angle=0]{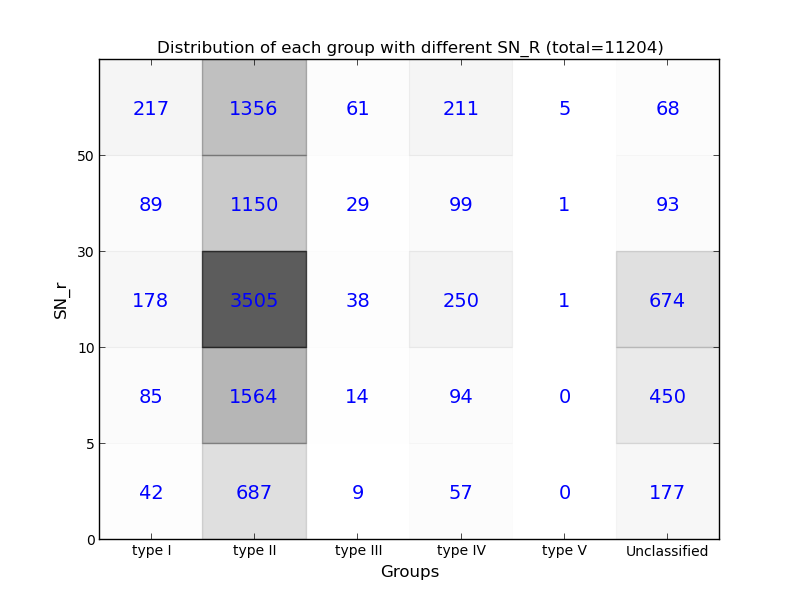}
  % \begin{minipage}[]{85mm}
   \caption{The statistics of six groups with different SN$\_$r. The x axis represents type I, type II, type III, type IV, type V, and the unclassified profile from left to right. The y axis represents five groups of different SN$\_$r, which are $<$5, 5$\sim$10, 10$\sim$30, 30$\sim$50 and $\geq$50 respectively. Each grid represents one type of SN$\_$r in a certain range, and the number of each group is shown by the gray level of its grid.} 
%\end{minipage}
   \label{Fig2}
   \end{figure}

Classification according to the shapes of H$\alpha$ profiles for each group with different SN$\_$r is made using the method of cross--correlation. The 27 templates used in the matching process contain five types of H$\alpha$ profile from single peak to P--Cygni with different intensity, which are visually chosen from spectra with SN$\_$r higher than 50 in our sample. For a more accurate result, spectra for each types after template matching are also inspected by eyes carefully. The statistical number of five groups corresponding to different categories as well as the group of unclassified profiles from low to high SN$\_$r is presented in Figure \ref{Fig2}. 
A detailed description of each type listed in this figure is given in the following, including the classification result and the characteristic of each type for LAMOST spectra. 

(a) \textit{single--peak profile (type I and type II):} In our sample, single--peak profiles comprise of type I and type II, which account for 5.45 $\%$ and 73.7 $\%$ respectively. Both types have emission lines with one peak, which are roughly symmetrical. The intensity of emission lines has a large span from tiny to large. Different from type I with only one emission line, type II is characterized by single--peaked emission component superimposed on a broad absorption component which originates from photosphere of stars.

(b) \textit{double--peak profile (type III and type IV):} Double--peak profiles also include two types. Profiles of type III and type IV are observed in 1.35$\%$ and 6.35$\%$ of 11,204 H$\alpha$ emission spectra. Similar to the first two types, the only difference is whether an additional broad underlying photospheric absorption exists in type III and type IV. These two types both contain two emission peaks separated by an absorption reversal which is approximately unshifted in most cases. The line profiles exhibit various structures, symmetric or asymmetrical. Furthermore, a few spectra with type IV have quite a sharp and deep absorption component in the center of double-peaked emission line, as spectra of Be shell stars.%"The line profiles exhibit various structures, symmetric or asymmetrical." is on the air.

(c) \textit{P--Cygni profile (type V):} A few spectra with P--Cygni profiles are also found in our sample. Out of 11 thousand spectra, only seven show evident P--Cygni profiles, which are categorized as type V. From the appearance of the spectra, P--Cygni profile has a blue absorption beyond the wing of the emission line. An accepted explanation of such P--Cygni profiles is that a gaseous envelope expands away from the star \citep{1998Ap&SS.260...63L,2007sks..book.....R}. 

(d) \textit{unclassified profile:} About 13.0 $\%$ of all the H$\alpha$ emission spectra are placed in the group with unclassified profiles. This group comprises of spectra with H$\alpha$ profile affected by noises or showing more complex shapes. Be limited to the resolving power of LAMOST instruments or disturbed by noise, an exact type can not be given accurately for the H$\alpha$ profile interfered by noise. Besides, some spectra display more complex shapes of H$\alpha$ profiles, such as the triple--peak profile which can be present in the spectra of nova. However, the profiles with more complex shapes can change over short periods. A detailed explanation of the formation can be found in \citet{2004A&A...426..669B}.

In summary, profiles of type II are the most common in early--type H$\alpha$ emission stars of our sample and account for the largest proportion of 73.7$\%$ seen from Figure \ref{Fig2}. Besides, a considerable number of spectra, about 13.0$\%$ percent of all the spectra can not be accurately classified as any of five categories. which is probably caused by the interference of noise on H$\alpha$ profiles or a more complexity of H$\alpha$ shapes. These spectra are considered as an individual unclassified group. Finally, we point out that the categories are just based on spectra with the resolution of $\sim$ 1800. Many of the single-- and double--peak H$\alpha$ profiles don't show micro structures limited to the low resolution of spectra.

% figures.

\section{Physical classification of the H$\alpha$ emission stars}
Various H$\alpha$ emission profiles are produced by different physical mechanisms for different types of objects. In general, the H$\alpha$ emission is always indicative of the existence of stellar envelopes, accretion or outflow activities in the early--type emission--line stars. It is known that the H$\alpha$ emission line originates from the outer disks for CBe and HAeBe stars, and from the binary interaction for close binary like Algol system\citep{2007ASSL..342.....K,2007sks..book.....R}. The emission lines excited from the mechanisms mentioned above are all produced from stars. Meanwhile, there is one situation that the H$\alpha$ emission in stellar spectrum may not be intrinsic, which can originate from diffuse nebula spreading over the interstellar space. We give a comprehensive analysis of our sample from the perspective of physical classification in the following.

\subsection{Objects archived by available catalogues}
In order to figure out the number of newly discovered objects, it is necessary to make sure how many known stars are recorded by the previous literatures in our sample. Objects with definite categories recorded by SIMBAD and other catalogues are discussed in this section\citep{1999AAS..134..255K, 2005NewA...10..325Z, 2008MNRAS.384.1277W, 2011AJ....142..149N, 2015AJ....149....7C, 2015RAA....15.1325L}.

By use of the coordinates of objects, the supplementary information of 1462 spectra are retrieved from SIMBAD on--line database, which satisfy a search radius of 3 arcsec. By scanning the matching results, we find that about 1300 stars are flagged as "star" or "emission--line star" without more detailed object type in SIMBAD database. Given classifications for the remaining stars cover a variety of object types. Among these objects, it is specifically noted that 7 CBe stars, 3 HAeBe stars and 31 close binaries are found in our sample. In addition, there are a small number of stars labelled as white dwarf or white dwarf candidates, supergiant stars or horizontal branch stars. In the classification of objects in our sample, the highest priority is given to the results from SIMBAD.

Besides, several catalogues related to CBe, HAeBe stars or H$\alpha$ emission stars have been published in recent years \citep{1999AAS..134..255K, 2005NewA...10..325Z, 2008MNRAS.384.1277W, 2011AJ....142..149N, 2015AJ....149....7C, 2015MNRAS.446..274R, 2015RAA....15.1325L}. We also cross--match our sample with these catalogues to pick out the known CBe and HAeBe stars. Meanwhile, there are 196 H$\alpha$ emission stars without a more detailed spectral type which have been recorded in other catalogues. Table \ref{cross-match} presents a total for the number of stars cross--matched to existing catalogues.

\begin{table}
\bc
\begin{minipage}[]{160mm}
\caption[]{The number of known stars by cross-matching our sample with six other catalogues. \label{cross-match}}\end{minipage}
\setlength{\tabcolsep}{8pt}
\small
 \begin{tabular}{ccc}
  \hline\noalign{\smallskip}
Object type & Number & Reference \\
  \hline\noalign{\smallskip}
CBe stars & 10 & \citet{2011AJ....142..149N} \\
CBe stars & 11 & \citet{2015AJ....149....7C} \\
CBe stars & 161 & \citet{2015RAA....15.1325L} \\
CBe stars & 4 & \citet{2005NewA...10..325Z}\\
H$\alpha$ emission stars & 74 &  \citet{1999AAS..134..255K} \\ 
H$\alpha$ emission stars & 122 & \citet{2008MNRAS.384.1277W} \\
  \noalign{\smallskip}\hline
\end{tabular}
\ec
\end{table}

\subsection{Spectra contaminated by \Ion{H}{II} regions}
Within our sample, there are some spectra which may be contaminated by \Ion{H}{II} regions. In such cases it is difficult to determine if the emission is intrinsic or extrinsic to the star itself. Therefore, we select all the stars whose coordinates are located in the \Ion{H}{II} regions as an individual group by cross-matching with the catalogue provided by \citeauthor{1976A&AS...25...25D} (as shown in Figure \ref{Figadd_1}).

There are some spectra that should be excluded from this group although they are projected into \Ion{H}{II} regions. One case is the stars recorded as emission--line stars by SIMBAD on--line catalogue. The other case is the spectra which exhibit ionized iron emission lines originating from stars rather than \Ion{H}{II} regions. Finally, 3,600 spectra are categorized as the group of spectra contaminated by \Ion{H}{II} regions.

\subsection{Close binaries}
The interaction among the close binaries is one of the formation mechanisms for the H$\alpha$ emission line. Since sufficient information can not be extracted from the spectra for identification, we obtain the close binaries only by cross-matching with SIMBAD on--line catalogue. SIMBAD lists 31 close binaries form our sample comprising 19 eclipsing binaries and 12 cataclysmic variable stars (2 dwarf nova, 3 nova, 2 nova--like stars and 5 without subtypes). 

\subsection{CBe and HAeBe stars}
Apart from ordinary stars and those stars whose spectra are contaminated by \Ion{H}{II} regions, the remainder are predominately CBe and HAeBe stars. In order to distinguish these two types of stars, \citet{1984A&AS...55..109F} proposed a good criterion of ($\emph{H}$-$\emph{K}$, $\emph{K}$-$\emph{L}$) diagram on the basis of a much stronger IR excesses of HAeBe than CBe stars. However, since observations on $\emph{L}$ band are not only lacking but also not deep enough, it is reasonable to replace $\emph{L}$ band by $\emph{W1}$, the first band of WISE \citep{2010AJ....140.1868W}, in the color--color diagram for a large dataset. The magnitudes of $\emph{H}$ and $\emph{K}$ are collected from 2MASS \citep{2006AJ....131.1163S} and UKIDSS \citep{2006MNRAS.372.1227D}. Altogether 6,330 out of 7,480 spectra simultaneously have $\emph{H}$, $\emph{K}$ and $\emph{W1}$ available.

\subsubsection{The analysis of photometric systems}
\textit{The system errors between 2MASS and UKIDSS.} Since the magnitudes of $\emph{H}$ and $\emph{K}$ bands are retrieved from 2MASS and UKIDSS, we make an analysis of the system errors between these two $\emph{J}\emph{H}\emph{K}$ systems. We compare the color ($\emph{H}$-$\emph{K}$) and the magnitude of $\emph{K}$ band between 2MASS and UKIDSS which are involved in the color--color diagram. Because the number of common objects in both UKIDSS and 2MASS is quite small for our sample, we obtain the magnitudes of UKIDSS transformed from 2MASS through an empirical formula derived by \citet{2006MNRAS.372.1227D}.  The residual of ($\emph{H}$-$\emph{K}$) between two systems is 0.02 mag which is comparable to the observational uncertainty of 2MASS, and the residual of K magnitude is 0.005 mag that can be negligible. Since these residuals have small effect on the classification of objects in the ($\emph{H}$-$\emph{K}$, $\emph{K}$-$\emph{L}$) diagram, the IR data from 2MASS in combination with UKIDSS is used without any calibrations.
%% figures.
%\begin{figure}
%   \centering
%   \includegraphics[width=12.0cm, angle=0]{Fig3.png}
%  % \begin{minipage}[]{85mm}
%   \caption{The comparisons between $\emph{J}\emph{H}\emph{K}$ systems of 2MASS and UKIDSS for ($\emph{H}$-$\emph{K}$) color and $\emph{K}$ magnitude are shown in the top panels. The histograms of residuals for ($\emph{H}$-$\emph{K}$) and $\emph{K}$ are shown in the bottom panel. The mean values and standard deviations of the residuals are 0.02 mag, 0.01 mag and 0.005 mag, 0.004 mag respectively for ($\emph{H}$-$\emph{K}$) and $\emph{K}$ band.} 
%%\end{minipage}
%   \label{Fig3}
%   \end{figure}

\noindent\textit{Replacing $\emph{L}$ with $\emph{W1}$ magnitude.} We also consider the effect of replacing $\emph{L}$ with $\emph{W1}$ on the color--color criteria. In order to analyse the system error between $\emph{L}$ and $\emph{W1}$ magnitudes, we use 184 objects measured $\emph{L}$ band magnitudes from \citet{2005A&A...438..663M} and \citet{2006A&A...450..253M}. A comparison reveals a mean difference of $\sim$~0.1 mag with a standard deviation of $\sim \pm$0.2 mag. The average reported observational accuracies for $\emph{L}$, $\emph{W1}$ magnitudes are 0.06 $\pm$ 0.08 mag, 0.02 $\pm$ 0.03 mag respectively. This result indicates that $\emph{W1}$ magnitude is basically consistent with $\emph{L}$ magnitude. However, due to the lack of observations for $\emph{L}$ band, the available samples are too few to make an accurate calibration between $\emph{L}$ and $\emph{W1}$ bands. Despite this lack of calibration, the strong coincidence and low scatter between the two bands give us confidence to replace $\emph{L}$ with $\emph{W1}$ in our color--color plot. Furthermore, we test the validity of ($\emph{K}$-$\emph{W1}$) color criteria by use of samples of known HAeBe and CBe stars. Using WISE $\emph{W1}$ band magnitude for 40 HAeBe stars with $\emph{L}$ magnitude \citep{1984A&AS...55..109F} and 122 CBe stars without $\emph{L}$ magnitude \citep{2011BASI...39..517M}, we plot the locations of these objects in the ($\emph{H}$-$\emph{K}$, $\emph{K}$-$\emph{W1}$) color--color diagram. The results show that only one HAeBe star and 13 CBe stars fall into the region in between. From this result, we conclude that the ($\emph{K}$-$\emph{W1}$) color--criteria can also separate HAeBe and CBe stars effectively although a small fraction of the sample scatter between the regions of CBe stars and HAeBe stars.

\subsubsection{interstellar extinction correction}
In the near IR region, the extinction due to interstellar medium along the line of sight to the star needs to be taken into account. Generally speaking, the interstellar extinction can be negligible for objects in high Galactic latitude areas (b $>30^{\circ}$). While for some objects in the low Galactic latitude areas especially in Galactic Anti-center, the extinction is significantly large. Therefore, the magnitudes of $\emph{H}$, $\emph{K}$ and $\emph{W1}$ of low Galactic latitude objects are supposed to be corrected before using the color--color criterion. In the correction, we adopt R$_V$ = 3.1 for A$_V$ = R$_V$ * E(B-V) \citep{1975A&A....43..133S, 1992ApJ...397..613H, 2006Ap&SS.305...11Z} and the extinction law A$_\lambda$/A$_V$ from \citet{2013MNRAS.430.2188Y} where values of A$_\lambda$/A$_V$ are 0.19, 0.13 and 0.082 for $\emph{H}$, $\emph{K}$ and $\emph{W1}$ band respectively.

The color excess E(B-V) of each object is obtained from two ways. 3,925 spectra in our sample have the value of E(B-V) with an accuracy of 0.04 mag retrieved from LAMOST value--added catalogues for LSS--GAC \citep{2015MNRAS.448..855Y,2015xiangms}. For objects not found from LAMOST value-added catalogues, E(B-V) is estimated from (B-V) - (B-V)$_{0}$. The observed color (B-V) is taken from NOMAD \citep{2004AAS...205.4815Z} and the intrinsic color (B-V)$_{0}$ of stars for each spectral type are taken from \citet{1970A&A.....4..234F}. Although the maximum error of Av is about 1 mag due to the uncertainty of spectral subtype determined by LAMOST pipeline, it is accepted for ($\emph{H}$-$\emph{K}$, $\emph{K}$-$\emph{W1}$) color--color classification \citep{1984A&AS...55..109F}. Finally, all the low Galactic latitude stars with the magnitudes of $\emph{H}$, $\emph{K}$ and $\emph{W1}$ available are corrected for interstellar extinction according to the methods mentioned above.
  
\subsubsection{Application of (H-K,K-W1) color--color criteria}
\begin{figure}
   \centering
   \includegraphics[width=12.0cm, angle=0]{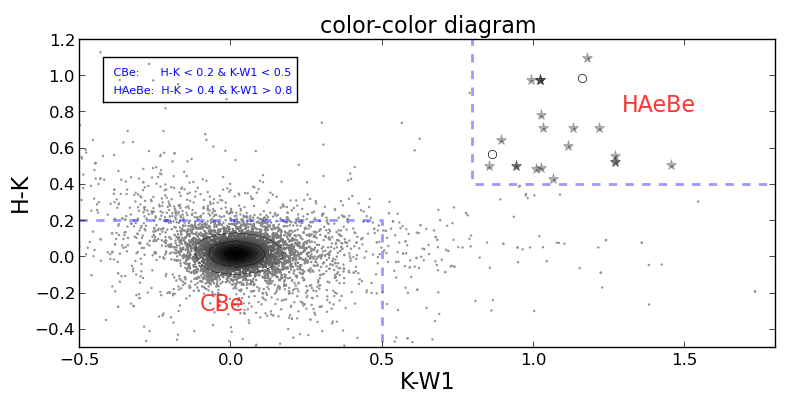}
  % \begin{minipage}[]{85mm}
   \caption{The $\emph{H}$-$\emph{K}$ vs $\emph{K}$-$\emph{W1}$ color--color diagram for 6,330 early--type H$\alpha$ emission stars from LAMOST DR2. The dotted lines represent the color criteria of CBe and HAeBe stars \citep{1984A&AS...55..109F}. 21 newly discovered HAeBe stars are plotted by the asterisks, and two known stars are marked by open circles. The darker asterisks indicate multiple observations.} 
%\end{minipage}
   \label{Figadd_2}
   \end{figure}

\citet{1984A&AS...55..109F} pointed out that HAeBe stars typically occur in the region where $\emph{H}$-$\emph{K}$ $>$ 0.4 and $\emph{K}$-$\emph{L}$ $>$ 0.8, while CBe stars have $\emph{H}$-$\emph{K}$ $<$ 0.2 and $\emph{K}$-$\emph{L}$ $<$ 0.5. Before using the color criteria, the extinction corrections have been made for objects with low Galactic latitude which is discussed above. Then we apply this criterion to our sample with $\emph{L}$ replaced by $\emph{W1}$. All the objects which have available photometric data, de-reddened if necessary, have been placed in the $\emph{H}$-$\emph{K}$, $\emph{K}$-$\emph{W1}$ color--color diagram, as shown in Figure \ref{Figadd_2}. As a result, the large dataset is divided into three groups containing 23 HAeBe stars, 5,594 CBe stars and 713 spectra in between. In 23 HAeBe stars, two objects have been investigated in the literature \citep{2010AJ....139...27S,2014IBVS.6089....1K}, which are marked by open circles in Figure \ref{Figadd_2}. For the objects scattering between the regions of CBe stars and HAeBe stars, we speculate that they are probably either CBe or HAeBe stars due to the uncertainty of IR color and the extinction correction. Additionally, it should be pointed out that B[e] stars may be also located in the region of HAeBe stars in the ($\emph{H}$-$\emph{K}$,$\emph{K}$-$\emph{W1}$) diagram, due to a large IR excess caused by hot circumstellar dust \citep{2006ASPC..355...13M}. Classification of these stars needs more information and further studies.

% figures.

\subsection{Analysis of the H$\alpha$ line profiles according to classifications}
\begin{figure}[!htbp]
   \centering
   \includegraphics[width=5cm]{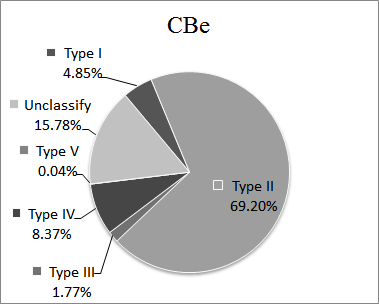}
   \includegraphics[width=5cm]{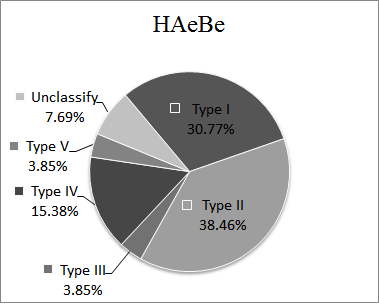}\\
   \includegraphics[width=5cm, angle=0]{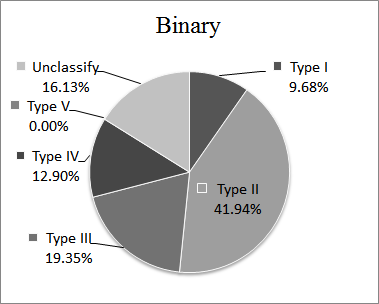}
   \includegraphics[width=5cm, angle=0]{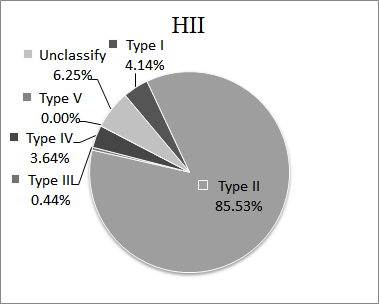}
  % \begin{minipage}[]{85mm}
   \caption{The proportions of different H$\alpha$ profile shapes for CBe stars, HAeBe stars, close binaries and spectra contaminated by \Ion{H}{II} regions for the low resolution spectra in our sample (R=1800).} 
%\end{minipage}
   \label{Fig9}
   \end{figure}

The mechanisms which give rise to H$\alpha$ emission in CBe stars, HAeBe stars, close binaries and stars within \Ion{H}{II} regions can vary according to physical processes. Moreover, for a certain type of stars, the shapes of H$\alpha$ emission profiles can be caused by various physical process. For investigating the distributions of different H$\alpha$ profile shapes in each group, we make a statistical analysis of the whole sample. The results are shown in Table \ref{table2} and Figure \ref{Fig9}.

For CBe stars, as seen from top--left panel of Figure \ref{Fig9}, the shapes of H$\alpha$ emission lines consist of single--peak, double--peak and P--Cygni profiles and they are dominated by type II (69.20$\%$). A traditional and competitive explanation for the diversity in the line profile shapes is the different angles of the rotational axis of the star with respect to the observer's line--of--sight. \citet{2007ASSL..342.....K} has given a general description of various line profiles which may arise due to the inclination angle, which indicates double--peak profiles are the most prevalent among the spectra of CBe stars.

Similarly, all five types of H$\alpha$ profiles are present in the spectra of HAeBe stars of our sample (the top--right panel of Figure \ref{Fig9}). A variety of theories have been proposed to interpret the H$\alpha$ profiles of different shapes. For P--Cygni profiles, it is widely accepted that they originate in the stellar wind. Other profiles have been explained by several models, such as clumpy circumstellar environment, magnetic models, winds with velocity gradients and rotation \citep{1994ASPC...62...91C,1996A&AS..120..229R}. From Figure \ref{Fig9}, we conclude that more than half of the profiles have single peak for the low resolution spectra in our sample (R=1800), which is different from the sample from \citeauthor{1984A&AS...55..109F}. \citet{1984A&AS...55..109F} made a comprehensive study of H$\alpha$ profiles of 57 HAeBe stars or candidates using high--resolution spectra. They came to an conclusion that the double--peak profiles have the largest proportion, followed by single--peak and P--Cygni profiles. We infer that the overestimation of single--peak profiles and underestimation of double--peak profiles in both CBe and HAeBe stars of our sample probably results from the low resolution of spectra.

For close binaries and objects contaminated by \Ion{H}{II} regions, the distribution of each type for H$\alpha$ profiles are also shown in the bottom panels of Figure \ref{Fig9}. No P--Cygni profiles (type V) are found in both groups. Additionally, it is evident that the majority of spectra contaminated by \Ion{H}{II} regions show single peak (the bottom--right panel of Figure \ref{Fig9}).

\begin{table}
\bc
\begin{minipage}[]{160mm}
\caption[]{The number of CBe, HAeBe stars, close binaries and objects contamnated by \Ion{H}{II} regions with different H$\alpha$ profile shapes. \label{table2}}\end{minipage}
\setlength{\tabcolsep}{8pt}
\small
 \begin{tabular}{cccccccc}
  \hline\noalign{\smallskip}
Type & I & II & III & IV & V  & unclassified & total\\
  \hline\noalign{\smallskip}
CBe & 272 & 3877 & 99 & 469 & 2 & 884 & 5603\\
HAeBe & 8 & 10 & 1 & 4 & 1 & 2 & 26\\
close binaries & 3 & 13 & 6 & 4 & 0 & 5 & 31\\
\Ion{H}{II} region & 149 & 3079 & 16 & 131 & 0 & 225 & 3600\\
  \noalign{\smallskip}\hline
\end{tabular}
\ec
%% place \tablecomments and \tablerefs below \end{center| and \end{center}:
%% you may leave the table-width parameter to editors or set to your actual size
\tablecomments{\textwidth}{\\1. Type I to V represent single--peak emission, single--peak emission in absorption, double--peak emission, double--peak emission in absorption, and P--Cygni profiles respectively.\\2. Objects recorded by available catalogues are also added to each group which are presented in this table.}
\end{table}

\subsection{Variability of H$\alpha$ profiles}
There are some objects having more than one observations in our sample. In order to investigate the variability of the H$\alpha$ emission profiles, we statistically analyse the frequency of multiple observations. The sample of H$\alpha$ emission stars includes 11,204 spectra of 10,436 unique objects. A total of 9,745 sources have been observed once, and 691 objects have repeated observations varying from 2 to 6 times. Then the equivalent widths of H$\alpha$ lines are calculated for these stars with more than one spectrum. The information is listed in an individual table which is available on line (http://dr2.lamost.org/doc/vac
). Meanwhile, we check all the spectra with repeated observations visually. It is found that $\sim$70\% of the spectra display a stable H$\alpha$ profile shape, and $\sim$30\% exhibit an obvious change in the profile. There are two modes of the H$\alpha$ variability for the spectra in our sample, one of which is the change in  emission line intensity and the other of which is the transformation between two shapes such as the change from single--peak to double--peak profiles. Here, we need to point out that the times of observations and the quality of spectra play a key role on the reliability analysis of the H$\alpha$ variability.

\section{Analysis of the subsamples}
\subsection{A subsample with iron emission lines}
In the sample of spectra with H$\alpha$ emission from LAMOST DR2, the majority only display Hydrogen emission, in most cases low order Balmer series such as H$\alpha$ or/and H$\beta$ emission line. Sometimes, besides Balmer emission, metal emission always iron lines, are also present in a small number of spectra. The study of \Ion{Fe}{II} emission lines, when measurable in Be stars, can contribute to understanding the physical properties of the circumstellar envelopes and the geometry of Be stars \citep{1992ApJS...81..335S, 1995A&AS..111..457B, 1994IAUS..162..265H, 2006A&A...460..821A}.

\begin{figure}
   \centering
   \includegraphics[width=14.0cm, angle=0]{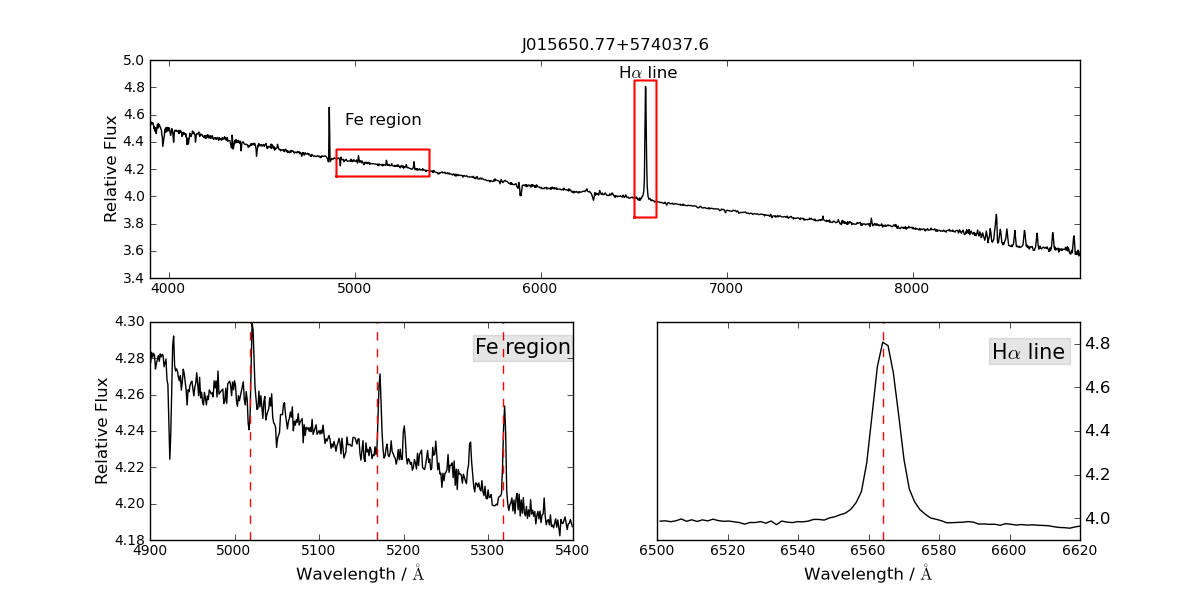}
  % \begin{minipage}[]{85mm}
   \caption{\Ion{Fe}{II} emission lines with single--peak. The spectrum of wavelength ranging from 3800 \AA~ to 8900\AA~ is plotted in the top panel. Two regions of the spectrum are shown in the bottom , which are the features of H$\alpha$ profile and iron lines respectively. Three lines of \Ion{Fe}{II} 5018 \AA, 5169 \AA, and 5317 \AA ~used in the emission detection are marked by the red dotted line in the left--bottom panel.} 
%\end{minipage}
   \label{Fig5}
   \end{figure}

In order to get such a subsample displaying both Balmer and metal emission, we have detected 3 iron lines of all the sample including \Ion{Fe}{II} 5018 \AA, 5169 \AA, and 5317 \AA. Due to a weak intensity of iron lines easily disturbed by noise in early--type spectra, a dataset with the signal to noise for g band higher than 10 (SN$\_$g$\geq$10) are chosen for \Ion{Fe}{II} emission detection. 172 spectra with the appearance of both intense iron and Balmer emission have been picked out from spectra with H$\alpha$ emission lines. The profiles of iron emission lines also have various types, including single--peak, double--peak(symmetric or asymmetric) and P--Cygni. Various types of metal line profiles are shown from Figure \ref{Fig5} to Figure \ref{Fig8}. Besides Balmer and iron features, many other common emission lines are also found in these spectra, such as \Ion{O}{I}, \Ion{Ca}{II} triplet, or Paschen series. A table of the 172 spectra with \Ion{Fe}{II} emission lines is provided on line (http://dr2.lamost.org/doc/vac), which includes four columns with the headings of designation, ra, dec and Fe$\_$type. The first column represents the designation in LAMOST, the second and the third one represent equatorial coordinates of the objects, and the last column represents the morphological type of \Ion{Fe}{II} emission lines.

% figures.
   
   \begin{figure}
   \centering
   \includegraphics[width=14.0cm, angle=0]{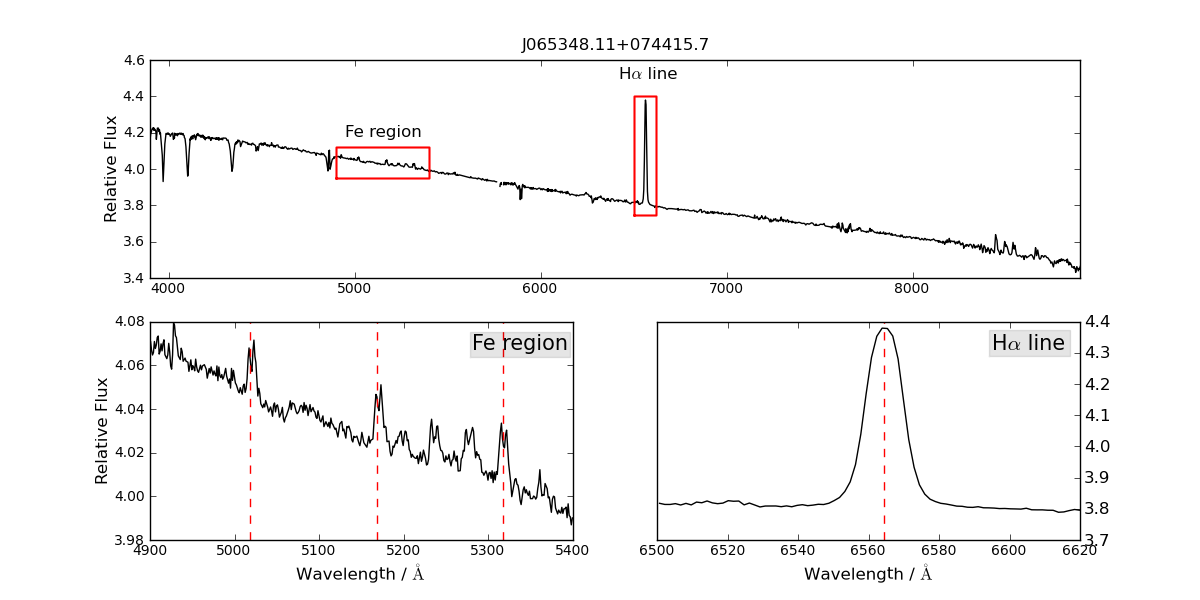}
  % \begin{minipage}[]{85mm}
   \caption{Same as Figure \ref{Fig5} but with \Ion{Fe}{II} emission lines showing symmetric double--peak.} 
%\end{minipage}
   \label{Fig7}
   \end{figure}
   
   \begin{figure}
   \centering
   \includegraphics[width=14.0cm, angle=0]{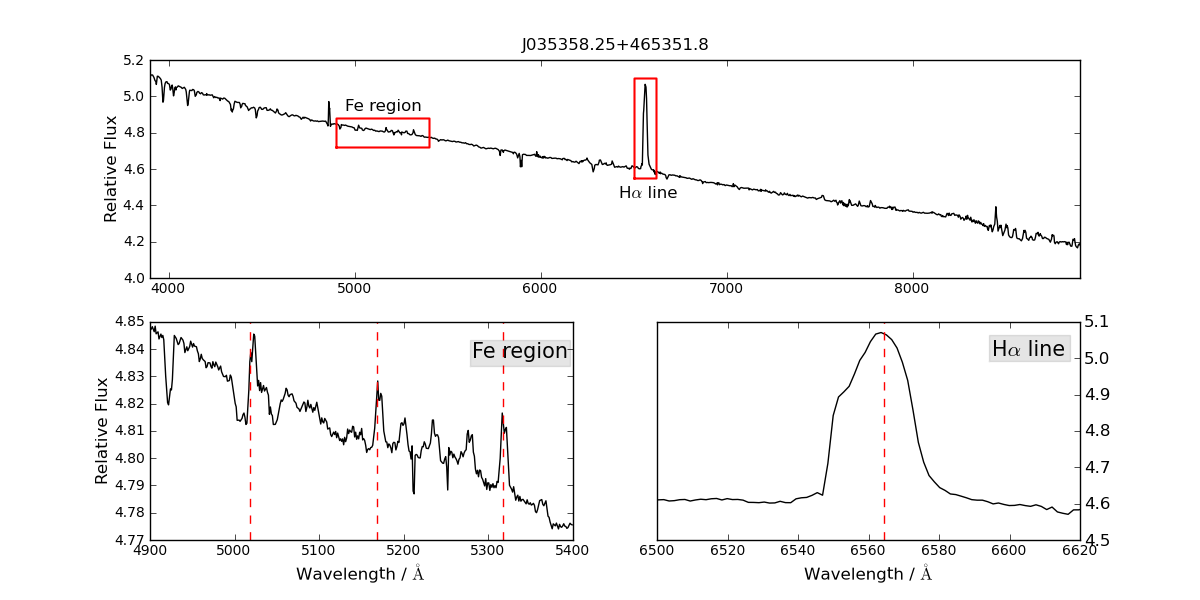}
  % \begin{minipage}[]{85mm}
   \caption{Same as Figure \ref{Fig5} but with \Ion{Fe}{II} emission lines showing asymmetric double--peak.} 
%\end{minipage}
   \label{Fig7}
   \end{figure}
   
   \begin{figure}
   \centering
   \includegraphics[width=14.0cm, angle=0]{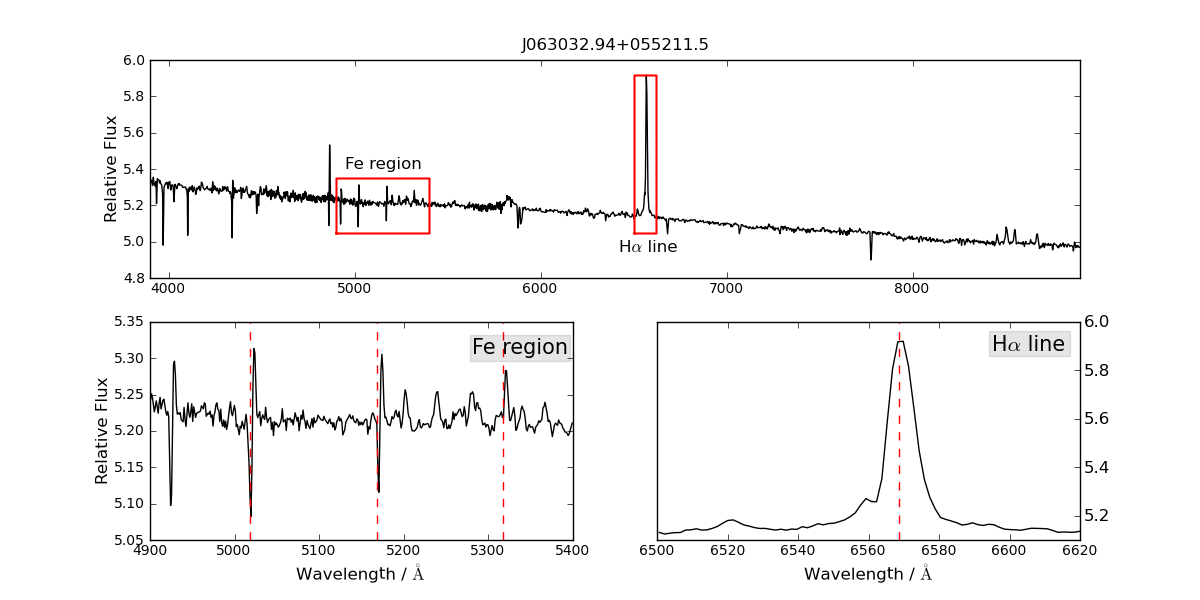}
  % \begin{minipage}[]{85mm}
   \caption{Same as Figure \ref{Fig5} but with \Ion{Fe}{II} emission lines showing P--Cygni profiles.} 
%\end{minipage}
   \label{Fig8}
   \end{figure}

\subsection{P--Cygni profiles and stellar wind velocity}
Seven spectra having H$\alpha$ emission with P--Cygni profile for five unique objects are discussed in this section. These profiles consist of a broad intense emission line with an absorption line displaced to the blue side of the emission component. It has been well known that such profiles indicate the existence of an outflowing wind in the neighbourhood of the star. A detailed explanation of line formation and related physical processes have been given by \citet{1998Ap&SS.260...63L} and \citet{2007sks..book.....R}. The basic information of the seven spectra is listed in Table \ref{table4} (The 3th and 4th ones belong to the same stars. So do the 6th and 7th). 

\begin{figure}
   \centering
   \includegraphics[width=14.0cm,height=7.0cm, angle=0]{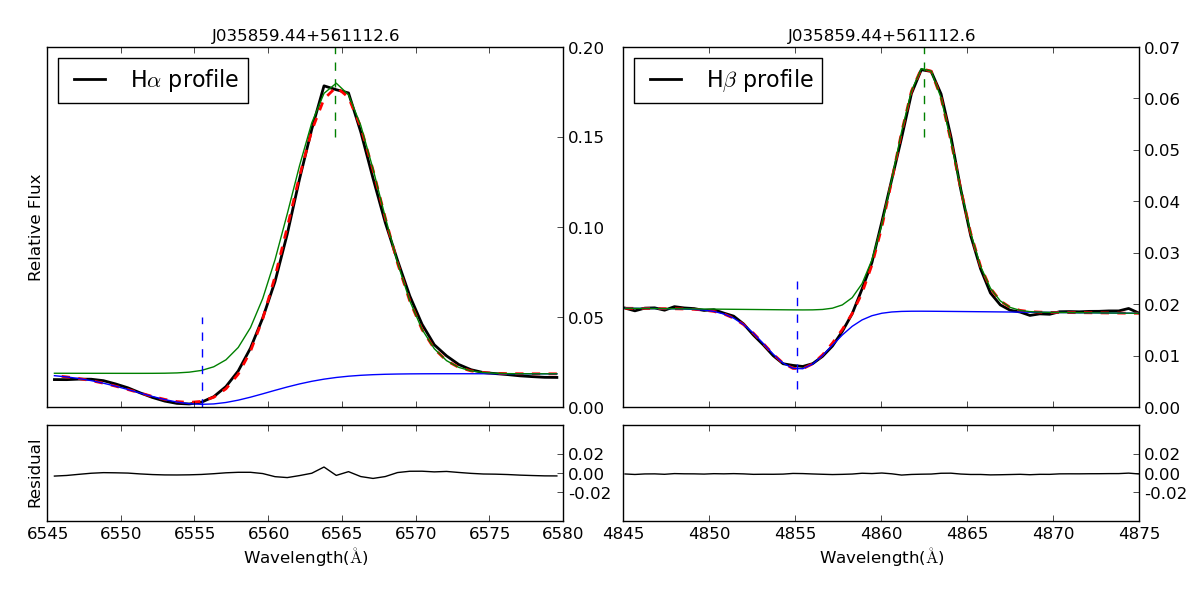}
  % \begin{minipage}[]{85mm}
   \caption{An example of the P--Cygni profiles of H$\alpha$ and H$\beta$ lines fitting by double--gaussian functions. In the top panels, the black line represents H$\alpha$ profile of the observational spectrum and the red dashed line is the result of double--gaussian fitting. The blue and green dotted lines represent the fitting results of absorption and emission component respectively. The residuals for the fitting is plotted in the bottom panels.} 
%\end{minipage}
   \label{Fig4}
\end{figure}

Specifically, the velocities of stellar wind for each object in the eighth column can be roughly derived from the Doppler effect formula: 
\begin{equation} \label{detection3}
 v_{wind} = (\bigtriangleup \lambda / \lambda_{0}) * c
\end{equation}

where $\bigtriangleup \lambda$ is the wavelength difference between the center of the emission and the absorption component, $\lambda_{0}$ is the center of the rest wavelength for the corresponding feature band, and c is the light velocity. In order to obtain the difference of wavelength center for two components ($\bigtriangleup \lambda$), we have fitted H$\alpha$ or/and H$\beta$ profiles of each spectra by the combination of two independent Gaussian profiles representing emission and absorption lines, as well as an linear function representing the local continuum. The fitting function is defined as follow.
\begin{equation} \label{detection4}
 f(x) = \sum_{i=1}^{2}A_{i} e^{-\frac{(x-\lambda_{i})^{2}}{2\sigma_{i}^{2}}}+(ax+b)
\end{equation}
where $A_{i}$, $\lambda_{i}$ and $\sigma_{i}$ represent the amplitude, the center and the standard deviation for the emission or absorption lines. Take H$\alpha$ profile as an example, $\bigtriangleup \lambda_{\alpha}$ (i.e. $\lambda_{1}$-$\lambda_{2}$) can be derived by the Gaussian fitting using Eq. \ref{detection4}. Then wind velocity $v_{\alpha}$ is calculated by simply plugging the difference of wavelength center $\bigtriangleup \lambda_{\alpha}$ into Eq. \ref{detection3}. Applying this method to H$\beta$ profile, we can also measure the wind velocity $v_{\beta}$ and the adopted wind velocity $v_{wind}$ is the average of $v_{\alpha}$ and $v_{\beta}$. An example of spectral lines with P--Cygni profile fitting by double Gaussian function is shown in Figure \ref{Fig4}.

% figures.

\begin{table}
\bc
\begin{minipage}[]{160mm}
\caption[]{The basic information of seven spectra with P--Cygni profiles. \label{table4}}\end{minipage}
\setlength{\tabcolsep}{4pt}
\footnotesize
 \begin{tabular}{cccccccc}
  \hline\noalign{\smallskip}
Designation ~~~~& Ra ~~~~& Dec ~~~~& $\lambda_{1blue}$ (\AA) & $\lambda_{1red}$ (\AA) & $\lambda_{2blue}$  (\AA) & $\lambda_{2red}$ (\AA) & V$_{wind}$ (km$\cdot$s$^{-1}$)\\
  \hline\noalign{\smallskip} 
J035859.44+561112.6 ~~~~& 59.747674 ~~~~& 56.186845 ~~~~& 6555.307108 & 6564.567527 & 4855.135384 & 	4862.518663 & 439.35\\
J063115.43+313054.5 ~~~~& 97.814293 ~~~~& 31.515164 ~~~~& 6558.924803 & 6564.867627 & / & / & 271.58\\
J063032.91+055217.8 ~~~~& 97.637126 ~~~~& 5.8716309 ~~~~& 6560.106444 & 6566.502805 & / & / & 292.31\\
J063032.94+055211.5 ~~~~& 97.637261 ~~~~& 5.869869 ~~~~& / & / & 4860.226369 & 4863.471923 & 200.23\\
J055054.77+201447.6 ~~~~& 87.72822 ~~~~& 20.246568 ~~~~& 6563.579882 & 6566.432534 & / & / & 130.37\\
J034753.05+291200.0 ~~~~& 56.971046 ~~~~& 29.200009 ~~~~& 6560.456203 & 6567.886835 & / & / & 339.58\\
J034753.05+291200.0 ~~~~& 56.971046 ~~~~& 29.200009 ~~~~& 6560.746404 & 6568.088851 & / & / & 335.55\\
  \noalign{\smallskip}\hline
\end{tabular}
\ec
%% place \tablecomments and \tablerefs below \end{center| and \end{center}:
%% you may leave the table-width parameter to editors or set to your actual size
\tablecomments{\textwidth}{\\1. Designation, Ra and Dec are retrieved from LAMOST catalogue.\\ 2. $\lambda_{1blue}$ and $\lambda_{1red}$ are the wavelength centers for the absorption and emission components obtained by a fitting for H$\alpha$ profiles; $\lambda_{2blue}$ and $\lambda_{2red}$ are for H$\beta$ profile. $`/'$ in the table indicates the emission lines without an evident P--Cygni profiles.\\
3. V$_{wind}$ in the last column are broad estimates for the stellar wind velocities.}
\end{table}

\section{Summary}
We provide a catalogue including 11,204 spectra for 10,436 early--type emission--line stars from LAMOST DR2, among which 9,752 early--type emission--line spectra are newly discovered. The catalogue can be accessed at the LAMOST Data Release web portal, http://dr2.lamost.org/doc/vac. It consists of twelve columns, which are respectively the designation in LAMOST, coordinates (Ra and Dec), $\emph{W1}$ magnitude and the error, $\emph{H}$ magnitude and the error, $\emph{K}$ and the error, morphological type of H$\alpha$ profiles, object types classified by our criteria and SIMBAD.

We also make a spectral analysis of the 11,204 spectra from LAMOST DR2. According to the H$\alpha$ line profiles, these spectra are classified into five distinct classes as well as a group of unclassified profiles. Moreover, four types of objects are found in our sample including CBe stars, HAeBe stars, close binaries and spectra contaminated by \Ion{H}{II} regions. Analyses of P--Cygni profiles and ionized iron emission lines are also included in this paper. Our analyses are based on a large sample of spectra with low resolution, and the sophisticated studies for interesting spectra need more information such as high--resolution spectra from follow--up observations. The data of LAMOST DR2 can be downloaded (http://dr2.lamost.org) to international researchers after the formal data release in June 2016.

% Authors can give a citation as `\citealt{Michel+etal+1992}'.
% You may also use \cite, \citep and \citet for citation, and use Table~1
% or Figure~1 and so forth. Using \ref and \label for cross-references of
% Tables/Figures is a good way in adjusting/adding/removing text, tables or

\normalem
\begin{acknowledgements}
We thank the anonymous referees for constructive comments, and BeSS. This work is supported by the National Key Basic Research Program of China (Grant No.2014CB845700), and the National Natural Science Foundation of China (Grant Nos 11390371, 11233004).

Guoshoujing Telescope (the Large Sky Area Multi-Object Fiber Spectroscopic Telescope, LAMOST) is a National Major Scientific Project built by the Chinese Academy of Sciences. Funding for the project has been provided by the National Development and Reform Commission. LAMOST is operated and managed by the National Astronomical Observatories, Chinese Academy of Sciences. In addition, we also thank BeSS catalogue provided by Neiner, C.
\end{acknowledgements}
  
\bibliographystyle{raa}
\bibliography{bibtex}

\end{document}